\documentclass[fleqn,twoside]{article}
\usepackage{espcrc2}
\usepackage{graphicx}

\title{6,13-dihydropentacene and pentacene single
co-crystals }\vspace{1pc}

\author{Christine C. Mattheus \address[grun]{Solid State Chemistry
          Laboratory, Materials Science Centre, University of Groningen, Nijenborgh 4,
          9747 AG Groningen, the Netherlands\\},
        Jakob Baas \addressmark[grun],
        Auke Meetsma \addressmark[grun],
        Jan L. de Boer \addressmark [grun],
        Christian Kloc \address[bell]{Bell Laboratories,
        Lucent Technologies, 600 Mountain Avenue, Murray Hill, NJ
        07974, USA},\\
        Theo Siegrist \addressmark[bell] \address[lund]{Department of Materials
        Chemistry, P.O. Box 124, Lund University, 221 00 Lund,
        Sweden}, and
        Thomas T. M. Palstra \addressmark[grun]\thanks{Corresponding author.
        Tel.: +31-50-363-4440; fax: +31-50-363-4441. E-mail:
        T.T.M.Palstra@chem.rug.nl} \vspace{0.5pc}}

\begin{document}

\begin{abstract}
\noindent {\bf{Abstract}}\\ 6,13-dihydropentacene and pentacene
co-crystallise in a ratio of 2:1 during vapour transport of
commercial pentacene in a gas flow. The crystal structure is
monoclinic P2$_1$/n and contains one dihydropentacene molecule,
and half a pentacene molecule in the asymmetric unit.\\

\noindent \it{Submitted to Acta Cryst. E. on 23 september 2002,
accepted for publication on 7 october 2002.\\ The crystallographic
information can be found at http://journals.iucr.org/e/ \\ or:
Mattheus Ph. D. Thesis, available for download at:
http://rugth30.phys.rug.nl/msc$\_$newweb/ph$\_$d.htm }

\vspace{1pc}
\end{abstract}

\maketitle

\section{Comment}
The growth of ultra-pure organic single crystals has recently
attracted much attention. Such crystals are prerequisite to
observe high electronic mobilities and band conduction in these
materials. We have studied single crystals of pentacene, which is
a material that, due to its high carrier mobilities and body of
published results, can be considered as model system to study
intermolecular interactions.

The vapour transport of commercially available pentacene powder
results in condensation of dark blue crystals of pentacene, either
as platelets, lath or as dendritic needles. During the vapour
transport growth, also crystals of 6,13-pentacenequinone
\cite{Dzyab} and 6,13-dihydropentacene-pentacene were obtained. We
have found that dihydropentacene co-crystallizes with pentacene to
form  long needle-like crystals. The colour of the crystals range
from dark pink/red to reddish to white transparent, with the same
structure. The 6,13-dihydropentacene molecules have two methylene
groups at opposite sides of the central ring. The co-crystal
structure is monoclinic, and crystallizes in space group P2$_1$/n.
The unit cell consists of two planar pentacene molecules and four
non-planar 6,13-dihydropentacene molecules, see Figure
\ref{fig:fig1}. The red needles were observed to grow along the
a-axis. 6,13-pentacenequinone and 6,13-dihydropentacene molecules
consist of two carbonyl groups and two methylene groups,
respectively, at each side of the central ring. This indicates
that the central ring of a pentacene molecule is the most reactive
site.

\begin{figure}[htb]
   \centering
   \includegraphics[bb= 20 640 250 810, width=60mm]{fig1small.eps}
   \caption{\emph{Unit cell of a 6,13-dihydropentacene and pentacene single
        co-crystal, viewed along the a-axis.}}
   \label{fig:fig1}
\end{figure}

\section{Experimental}
Two different crystal growth methods were used and both yielded
the same 6,13-dihydropentacene-pentacene co-crystals. As source
material for both methods pentacene powder (Aldrich) was used as
supplied. X-ray powder diffraction spectra of this commercially
available pentacene did not reveal any presence of contaminants,
such as pentacenequinone or dihydropentacene. However, small
amounts of contaminations cannot be detected by this method.
Method I (Groningen): single crystals were grown using physical
vapour transport in a horizontal glass tube \cite{Laudise}. A
pyrex glass tube, inner diameter of 16 mm, was cleaned by heating
under a nitrogen gas flow and placed in a second tube. 200-400 mg
of the source material were placed at the end of the tube in a
platinum crucible. The growth was either performed under a stream
of nitrogen gas mixed with hydrogen gas or under a pure argon
flow. The nitrogen and argon gasses were purified over activated
copper and alumina columns, to remove any residual traces of
H$_2$O and oxygen. Gasses were obtained from AGA, with 5N purity
for nitrogen and argon, and 4N5 for hydrogen. A temperature
gradient was applied by resistive heating of two heater coils
around the tube. The source material was sublimed at $\sim$ 550 K.
Depending on the exact temperature gradient, pentacene crystals
condensed at $\sim$ 300 mm from the sublimation point at a
temperature of $\sim$ 490 K. If the vapour transport is executed
in a poorly sealed system or residual oxygen as e.g. H$_2$O is
present 6,13-pentacenequinone crystals are observed at a slightly
higher temperature, $\sim$ 520 K.  However, if hydrogen is present
in the carrier gas either by dilution of by decomposition of the
starting material, 6,13-dihydropentacene-pentacene co-crystals
condense at lower temperature, \\sim 480 K. Increasing the amount
of hydrogen in the carrier gas was observed to increase the amount
of red crystals. The use of pure argon as transport gas yielded
significantly less red crystals.

Method II (Bell Labs): single crystals of pentacene were grown by
physical vapor phase transport in a horizontal transparent
furnace. A charge of a 10-30 milligrams of pentacene was placed in
a high temperature (280 -320 $^\circ$C) zone inside a two zone
furnace and was exposed to either an Ar, He or H$_2$O gas stream,
flowing at a rate of 40-100 ml min$^{-1}$. The pure gas, at a
pressure of $\it{ca.}$ 1 atm, was delivered to one end of the
crystal-growth reactor and exited from the system through a
bubbler, thus removing impurities and decomposition products. If
the impurities in pentacene consist of larger and/or smaller
molecules with either lower or higher vapor pressure, gas flow
will transport the smaller molecules from the hot zone to the cold
zone of the furnace. Molecules with higher vapor pressure will
remain in the hot zone and will not contaminate the growing
pentacene crystals. For molecules with very similar vapor
pressures the gas transport mechanisms are expected to be alike.
Pentacene crystals nucleated spontaneously on the reactor wall at
the low temperature region. The temperature of the center part of
the cold zone was set to 220 $^\circ$C, as measured by an external
sensor in close proximity to the heater. Due to flow of hot gas,
the temperature in the crystallization zone was spread from 320
$^\circ$C to room temperature over the length of the tube.
6,13-dihydropentacene-pentacene co-crystals were formed at the far
end of the furnace, where the temperature of the reactor tube
drops down to ambient temperature.

\begin{figure}[htb]
   \centering
   \includegraphics[bb= 30 675 275 820, width=70mm]{fig2rot.eps}
   \caption{\emph{View of the 6,13-dihydropentacene molecule and a pentacene
        molecule, showing the atom-labelling scheme. Displacement
        ellipsoids are drawn at the 50$\%$ probability level and H-atoms
        are shown as spheres of arbitrary radius.
        Symmetry code: (i) -x,-y,-z. }}
   \label{fig:fig2}
\end{figure}

\section{Refinement}
The X-ray structure determination was thwarted by persistent very
weak scattering of the crystals. After many trials a crystal  for
the structure determination was obtained, but no observed
reflections could be measured with $\theta > 20 ^\circ$. Due to
the low observation to parameter ratio the H atoms, which could be
located from difference Fourier maps, were included in calculated
positions and refined in the riding mode.

\begin{table}[htb]
   \centering
   \begin{tabular}{l|l}
     Formula&  2(C$_{22}$H$_{16}$)(C$_{22}$H$_{14}$) \\
     Formula weight&  839.03 g mol$^{-1}$  \\
     Crystal system&  monoclinic \\
     Space group, no. &  P21/n, 14 \\
     $a$&  6.163(4) \AA \\
     $b$&  21.801(5) \AA \\
     $c$&  16.069(3) \AA \\
     $\beta$ &  93.73(3)$^\circ$ \\
     Volume  &  2154.5(15) \AA$^{3}$ \\
     Formula Z&  2 \\
     Space group Z&  4 \\
     $\rho_{calc}$&  1.293 g cm$^{-3}$\\
     F(000)&  884 electrons \\
     $\mu$(Mo K$\alpha$)&  0.73 cm$^{-1}$  \\
     Colour, habit&  red, needle \\
     Approx. crystal dimensions&  $0.05\times0.07\times0.32$ mm$^3$ \\
      \\
   \end{tabular}
   \caption{\emph{Crystal data and details of the structure
   determination of 6,13-dihydropentacene and pentacene
   co-crystals}}
   \label{tab:dihydrocrystal}
\end{table}

\end{document}